\documentclass[fleqn,oneside]{article} % later: twoside
\usepackage{espcrc2}

\usepackage{epsfig} %\usepackage{graphicx}
\title{\vspace*{-10pt}%
Physics prospects of UV-filtered overlap quarks
}

\author{
S.\ D\"urr\address[UB]{Institut f\"ur theoretische Physik,
Universit\"at Bern, Sidlerstr.\,5, 3012\,Bern, Switzerland\vspace{-2mm}},
Ch.\ Hoelbling\address[UW]{Fachbereich Physik, Universit\"at Wuppertal,
Gaussstr.\,20, 42119\,Wuppertal, Germany\vspace{-2mm}} and
U.\ Wenger\address[ETHZ]{Institut f\"ur theoretische Physik,
Eidgen\"ossische Technische Hochschule, CH-8093 Z\"urich, Switzerland}
}
\begin{document}

\begin{abstract}
Some key features of the overlap operator with a UV-filtered Wilson kernel
are discussed. The first part concerns spectral properties of the underlying
shifted hermitean Wilson operator and the relation to the observed speedup of
the overlap construction. Next, the localization of the filtered overlap and
its axial-vector renormalization constant are discussed. Finally, results of
an exploratory scaling study for $m_{ud}, m_s$ and $f_\pi, f_K$ are presented.
%\vspace{-1pt}
\end{abstract}

\maketitle

\newcommand{\pad}{\partial}
\newcommand{\pas}{\partial\!\!\!/}
\newcommand{\Dsl}{D\!\!\!\!/\,}
\newcommand{\Psl}{P\!\!\!\!/\;\!}
\newcommand{\hqu}{\hbar}
\newcommand{\ovr}{\over}
\newcommand{\til}{\tilde}
\newcommand{\pri}{^\prime}
\renewcommand{\dag}{^\dagger}
\newcommand{\<}{\langle}
\renewcommand{\>}{\rangle}
\newcommand{\gaf}{\gamma_5}
\newcommand{\lap}{\triangle}
\newcommand{\trc}{{\rm tr}}
\newcommand{\nab}{\nabla}

\newcommand{\al}{\alpha}
\newcommand{\be}{\beta}
\newcommand{\ga}{\gamma}
\newcommand{\de}{\delta}
\newcommand{\ep}{\epsilon}
\newcommand{\ve}{\varepsilon}
\newcommand{\ze}{\zeta}
\newcommand{\et}{\eta}
\renewcommand{\th}{\theta}
\newcommand{\vt}{\vartheta}
\newcommand{\io}{\iota}
\newcommand{\ka}{\kappa}
\newcommand{\la}{\lambda}
\newcommand{\rh}{\rho}
\newcommand{\vr}{\varrho}
\newcommand{\si}{\sigma}
\newcommand{\ta}{\tau}
\newcommand{\ph}{\phi}
\newcommand{\vp}{\varphi}
\newcommand{\ch}{\chi}
\newcommand{\ps}{\psi}
\newcommand{\om}{\omega}
\newcommand{\psb}{\overline{\psi}}
\newcommand{\etb}{\overline{\eta}}
\newcommand{\psd}{\psi^{\dagger}}
\newcommand{\etd}{\eta^{\dagger}}
\newcommand{\beq}{\begin{equation}}

\newcommand{\eeq}{\end{equation}}
\newcommand{\bdm}{\begin{displaymath}}
\newcommand{\edm}{\end{displaymath}}
\newcommand{\bea}{\begin{eqnarray}}
\newcommand{\eea}{\end{eqnarray}}

\newcommand{\mr}{\mathrm}
\newcommand{\mb}{\mathbf}
\newcommand{\Nf}{{N_{\!f}}}
\newcommand{\Nc}{{N_{\!c}}}
\newcommand{\ri}{\mr{i}}
\newcommand{\DW}{D_\mr{W}}
\newcommand{\DWr}{D_{\mr{W},-\rh}}
\newcommand{\HWr}{H_{\mr{W},-\rh}}
\newcommand{\Dov}{D_\mr{ov}}

\newcommand{\Mpi}{M_\pi}
\newcommand{\Fpi}{F_\pi}
\newcommand{\MeV}{\,\mr{MeV}}
\newcommand{\GeV}{\,\mr{GeV}}
\newcommand{\fm}{\,\mr{fm}}

\hyphenation{topo-lo-gi-cal simu-la-tion theo-re-ti-cal mini-mum
re-nor-mali-za-tion}

%%%%%%%%%%%%%%%%%%%%%%%%%%%%%%%%%%%%%%%%%%%%%%%%%%%%%%%%%%%%%%%%%%%%%%%%%%%%%%

\section{INTRODUCTION}

On the conceptual level the quest for exact chiral symmetry at finite lattice
spacing has been completed.
The closely related domain-wall (in the limit $L_5\!\to\!\infty$)
\cite{domainwall} and overlap \cite{overlap} approach and also the parametrized
fixed point action (in the limit of perfect parametrization) \cite{fixedpoint}
yield lattice fermions which satisfy the Ginsparg-Wilson relation with on-shell
chiral symmetry \cite{Ginsparg:1981bj}
\beq
\gaf D+D\hat\gaf=0\;,\qquad\hat\gaf=\gaf(1-{1\over\rho}D)
\;.
\label{GW}
\eeq
The salient features of such fermions include:
\begin{itemize}
\itemsep-2pt
\item
Additive mass renormalization is absent \cite{Ginsparg:1981bj}.
\item
There is an index theorem linking $n_+-n_-$ (where $n_\pm$ is the number
of zero-modes of the massless Dirac operator $D$ with $\gaf\!=\!\pm1$) to the
gluonic topological charge $q$ on fine enough lattices \cite{indextheorem}.
\item
% The chiral charge is not quantized \cite{Luscher:1998pq}.
The\,Nielsen-Ninomiya\,theorem\,is\,evaded\,\cite{Luscher:1998pq}.
% There is a continuous symmetry \cite{Luscher:1998pq}.
\item
Using ``chirally rotated'' quark fields to construct e.g.\ pseudoscalar
densities and axialvector currents (see below) the theory has only $O(a^2)$
cut-off effects \cite{Capitani:1999uz}.
\end{itemize}

On a practical level the question how one may implement any such scheme at
bearable cost in terms of CPU time remains a topic of active research.
Considering the massless overlap operator
\beq
D_\mr{ov}=\rho\big[1+\DWr\big(\DWr\dag \DWr)^{-1/2}\big]
\label{def_overzero}
\eeq
with $\DWr\!=\!\DW\!-\!\rh$ the shifted Wilson operator ($0\!<\!\rh\!<\!2$)
as an example, two avenues are being pursued.
One question is what is the numerically most efficient way to implement the
$(.)^{-1/2}$ operator prescription in (\ref{def_overzero}).
%On the other hand one may ask whether a slight modification of the kernel
%in (\ref{def_overzero}) can reduce the computational burden.
%The filtering strategy discussed below belongs to this second category.
The second strategy is to ask whether a slight modification of the formulation
can reduce the computational burden.

%%%%%%%%%%%%%%%%%%%%%%%%%%%%%%%%%%%%%%%%%%%%%%%%%%%%%%%%%%%%%%%%%%%%%%%%%%%%%%%

\section{OVERLAP WITH UV-FILTERING}

An obvious idea is to do some ``massage'' to the Wilson kernel in the overlap
prescription, since there is nothing specific to the (plain) $\DW$ in
(\ref{def_overzero}); any legal doubler-free fermion action will do fine.

There is a long history of ``designer actions'' tailored to have a spectrum
sufficiently close to the GW circle, such that the inverse squareroot in
(\ref{def_overzero}) would minimize the number of forward applications of the
kernel on a source vector \cite{designer}.
These actions typically invoke tunable parameters and the improvement is
achieved through additional couplings, i.e.\ a single row or column of $D$
involves more than the $51$ %$3(16\!+\!1)\!=\!51$
entries (for $SU(3)$ and chiral repr.) of $\DW$.
Hence the challenge is to assess the effect of fewer forward applications
versus each application getting more expensive.

In addition%
\footnote{Some of the ``designer actions'' do involve smeared links.},
there is quite some experience with filtered actions (including the overlap)
\cite{oldfilt_stag,Hasenfratz:2001hp,oldfilt_wils,oldfilt_over} where only
the covariant derivative is replaced,
\beq
U_\mu(x)\ps(x\!+\!\hat\mu)\!-\!\ps(x) \to
V_\mu(x)\ps(x\!+\!\hat\mu)\!-\!\ps(x)
\;,
\label{def_filter}
\eeq
tantamount to a change by an irrelevant operator.
Here, $V_\mu(x)$ is defined via an APE \cite{Albanese:ds}, HYP
\cite{Hasenfratz:2001hp} or stout-link \cite{Morningstar:2003gk} recipe%
\footnote{The latter choice is mandatory in a dynamical HMC.}.
The point is that this maintains the sparseness of the Wilson operator
and yet a significant speedup can be achieved.
The obvious concern is that the filtering might impair the localization
properties of the final $D$.

\begin{figure*}
\vspace{-4mm}
\hspace{8mm}
\epsfig{file=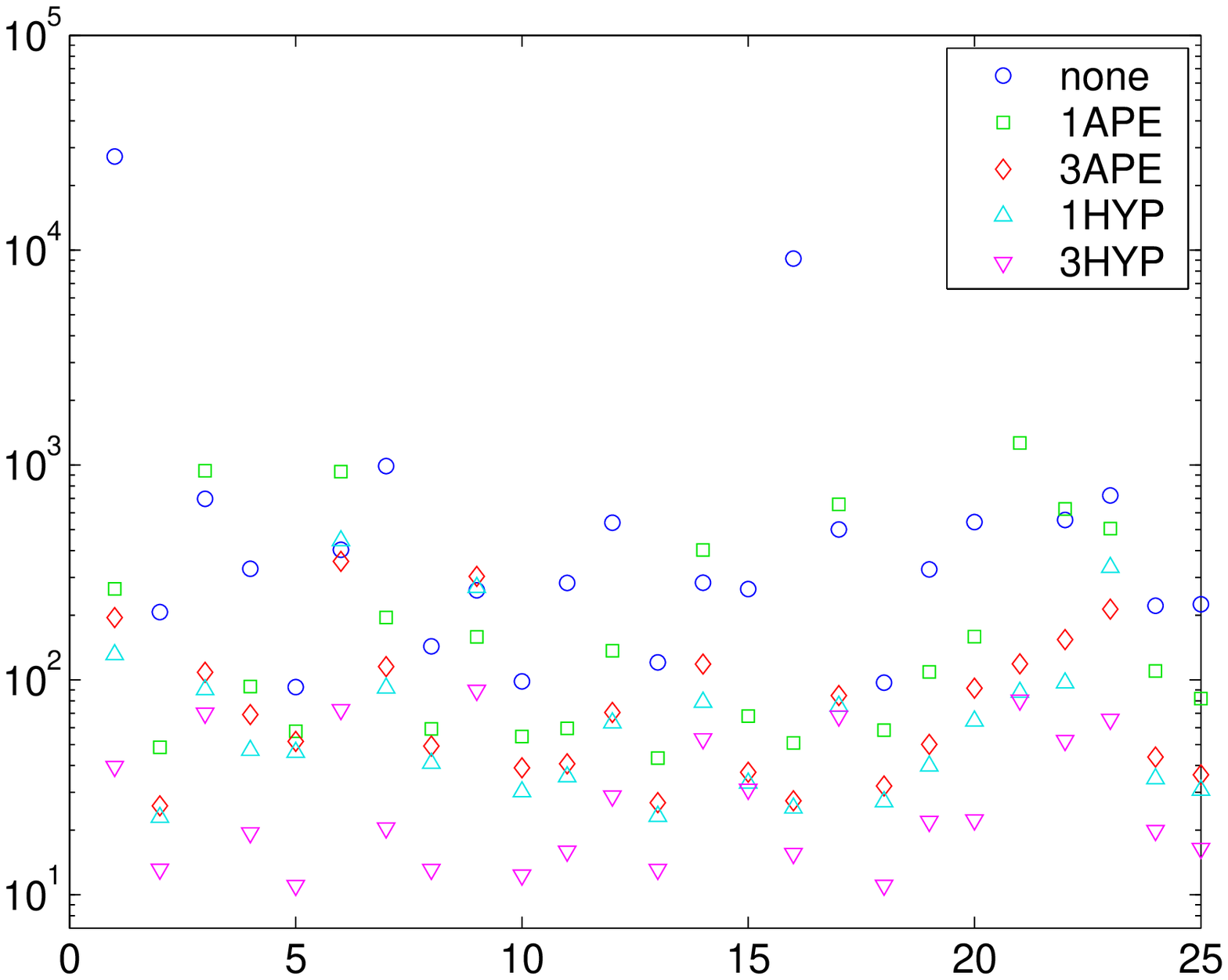,height=5.6cm} %6.4cm
\epsfig{file=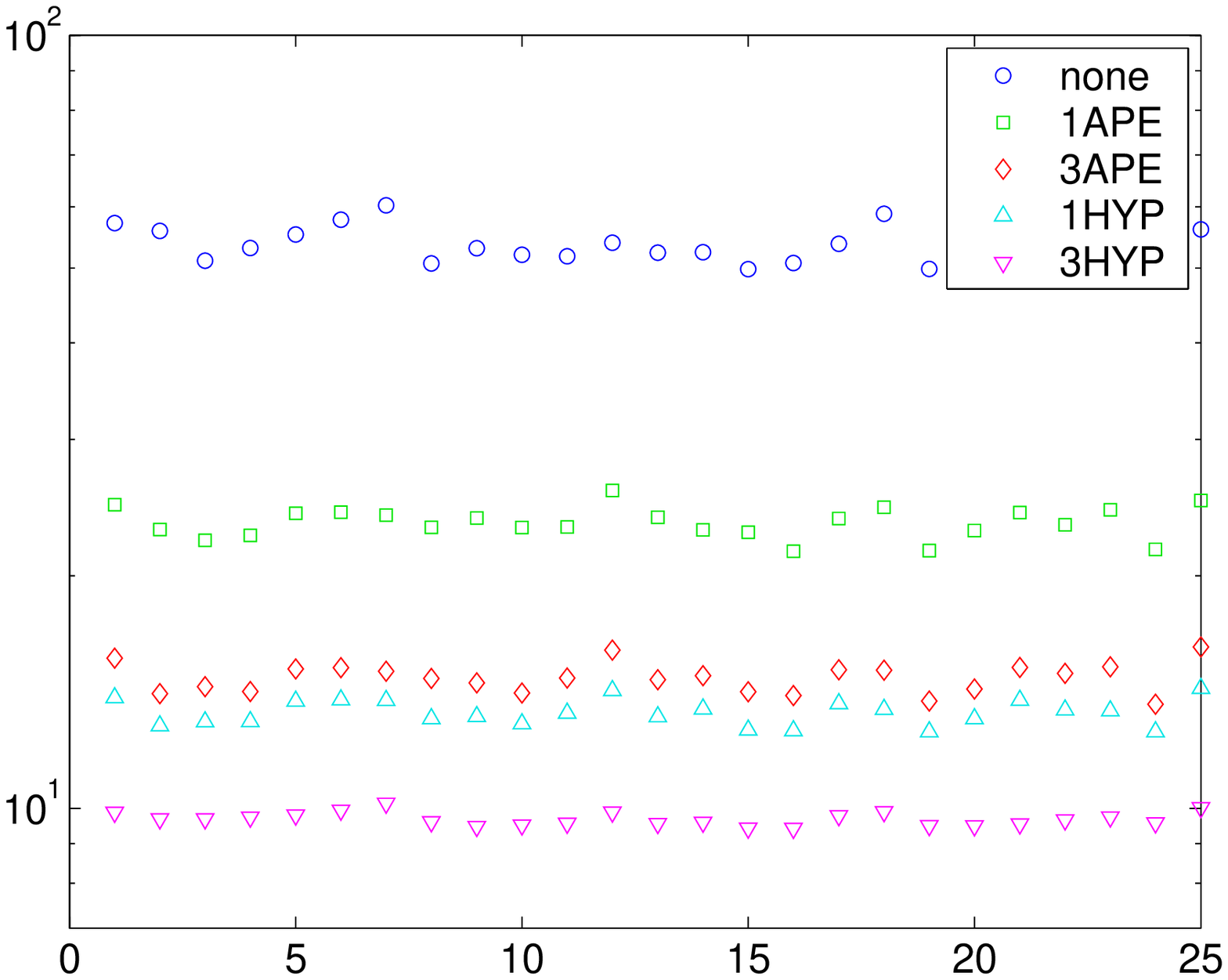,height=5.6cm} %6.4cm
\vspace{-10mm}
\caption{Condition number $1/\ep$ of $|\HWr|$ at $\rh\!=\!1$ on 25 quenched
$\be\!=\!6.0$, $16^4$ configurations without (left) and with projection of the
14 lowest eigenmodes (right). Modest filtering significantly reduces $1/\ep$.}
\vspace{2mm}
\hspace{8mm}
\epsfig{file=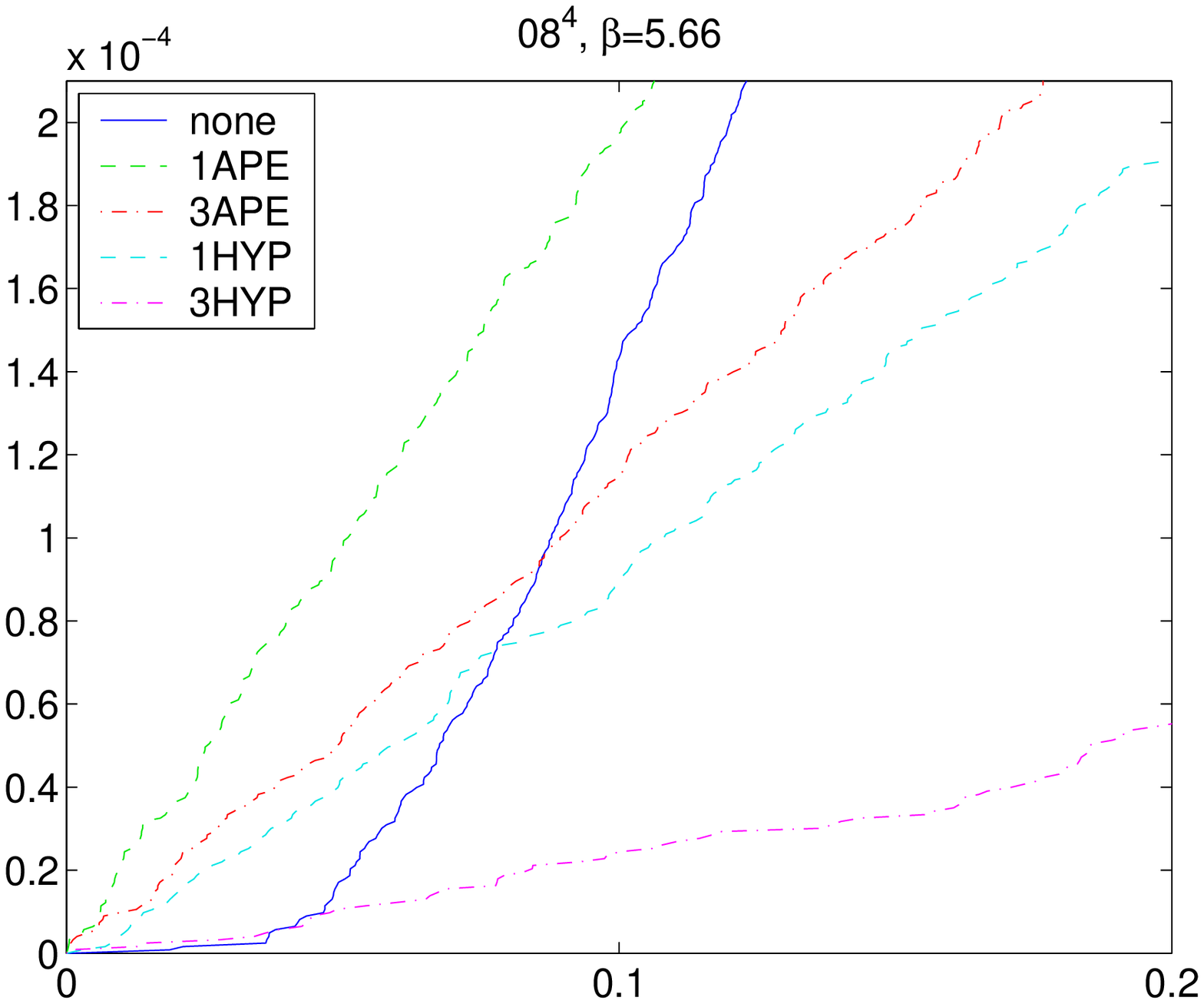,height=5.9cm} %6.7cm
\epsfig{file=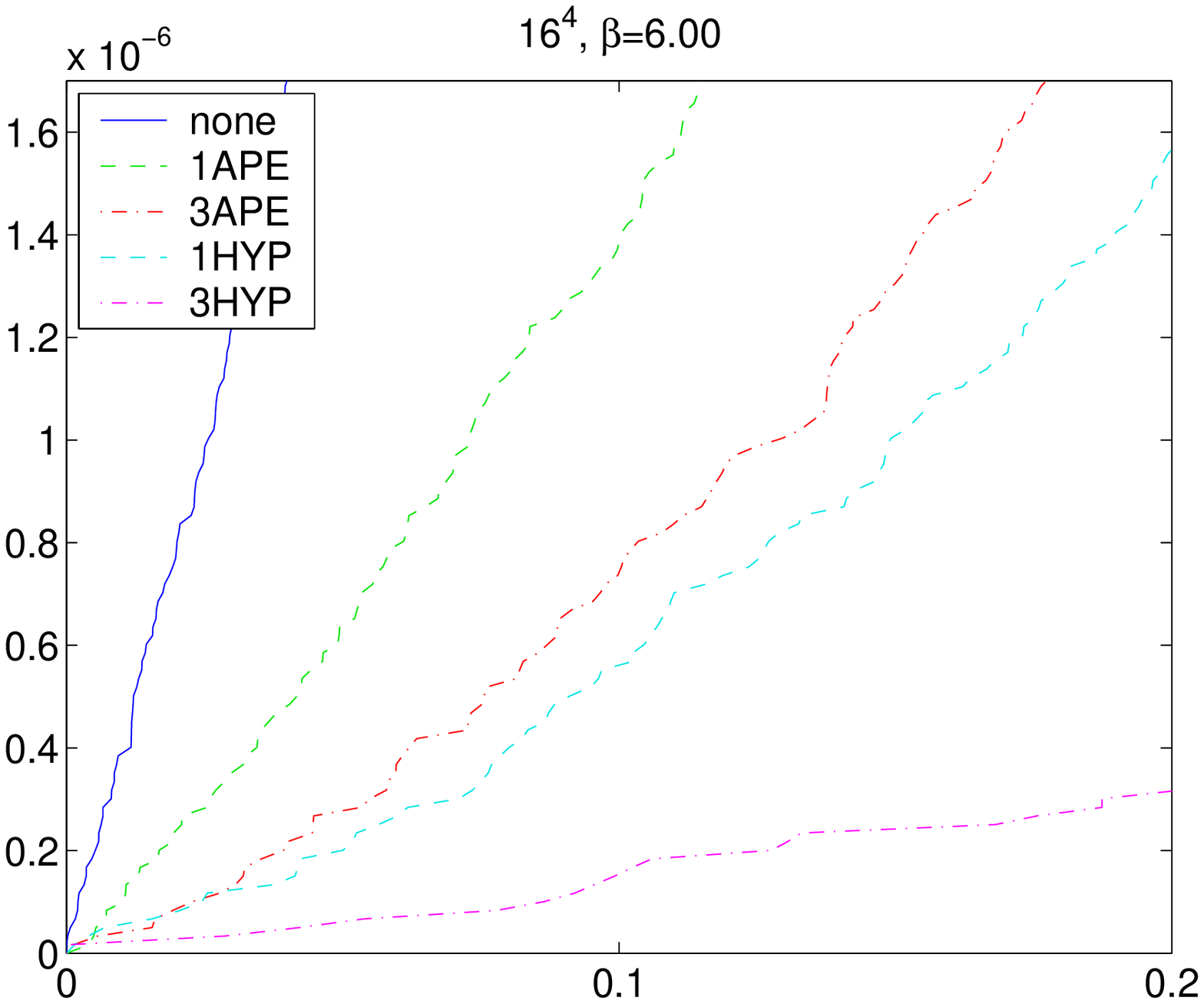,height=5.9cm} %6.7cm
\vspace{-10mm}
\caption{Cumulative eigenvalue distribution of $|\HWr|$ at $\rh\!=\!1$ for
$\be\!=\!5.66$ (left) and $\be\!=\!6.00$ (right) with 0,1,3 steps of APE or HYP
filtering. This reduces the slope and thus $\rh(0)$, but the latter quantity
remains non-zero. Evidently, the unfiltered operator at $\be\!=\!5.66$ is in
the wrong universality class.}
\vspace{2mm}
\hspace{8mm}
\epsfig{file=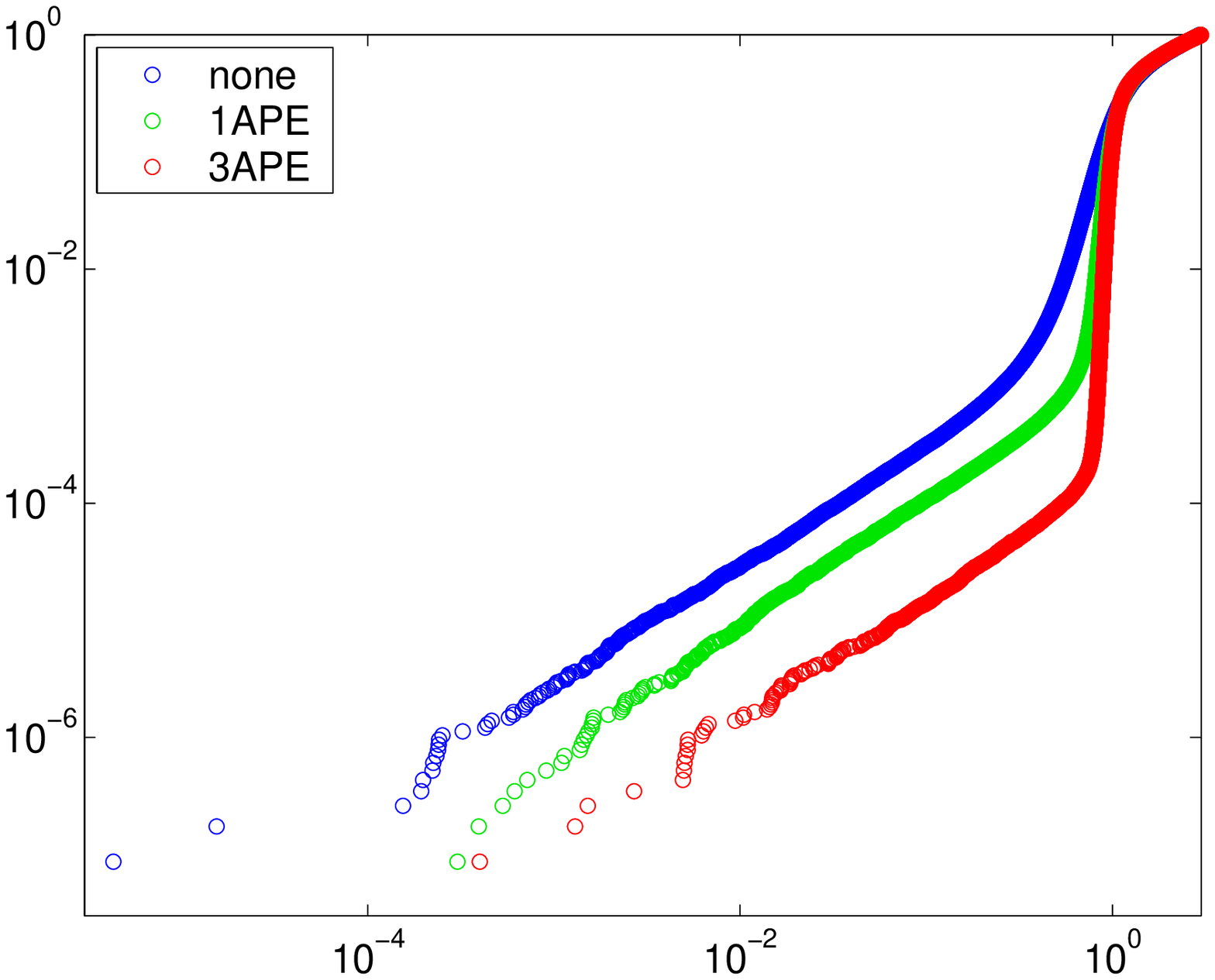,      height=5.6cm} %6.4cm
\epsfig{file=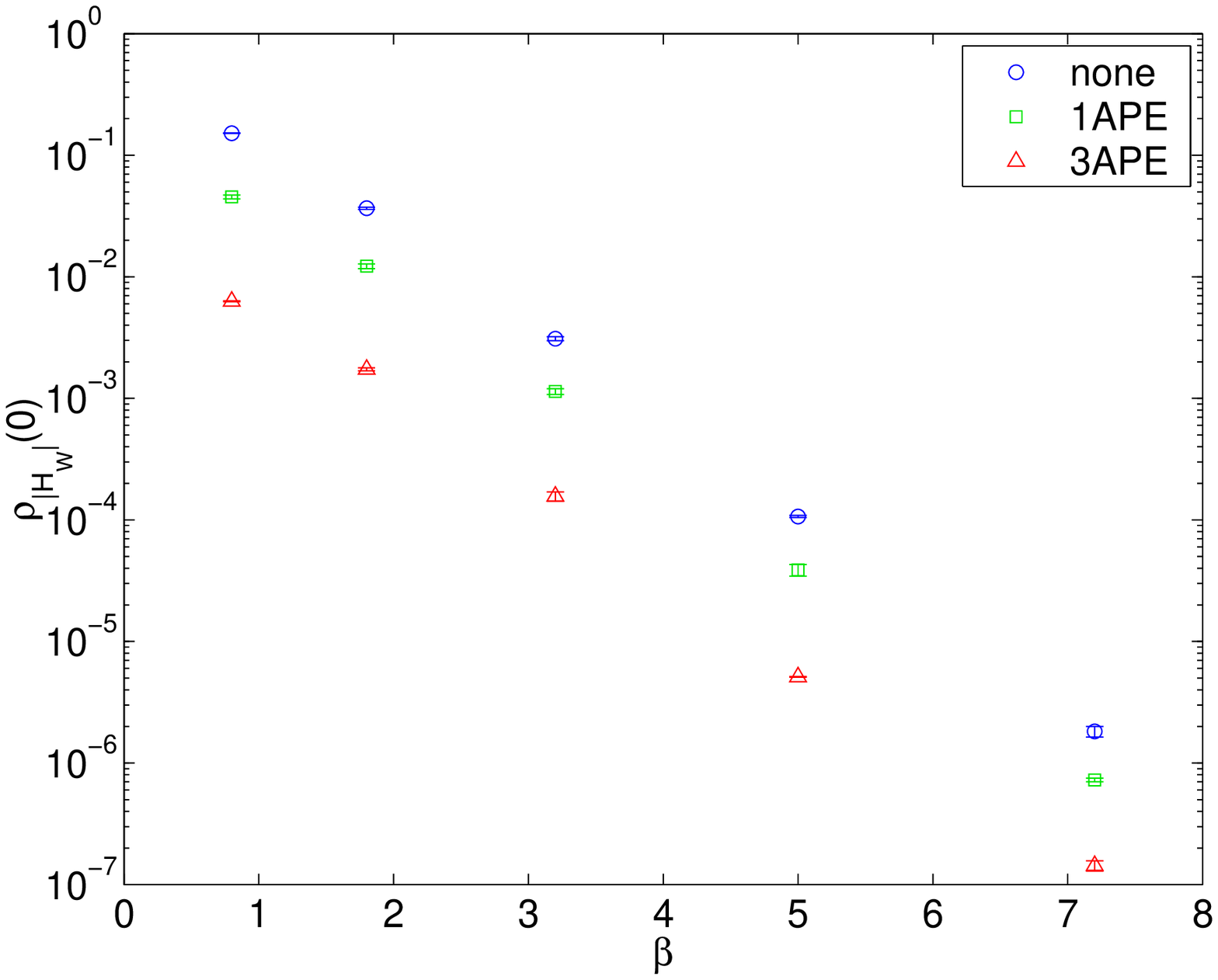,height=5.6cm} %6.4cm
\vspace{-10mm}
\caption{Cumulative eigenvalue distribution (log-log plot) of $|\HWr|$ in the
Schwinger model ($\Nf\!=\!0$, $16^2$, $\be\!=\!3.2$, $\rh\!=\!1$) with 0,1,3
filtering steps (left) and the pertinent $\rh(0)$ as a function of $\be$
(right).}
\end{figure*}

\begin{figure*}
\vspace{-4mm}
\hspace{8mm}
\epsfig{file=0506027.figs/eff_loc_L16_b6.00_r20.eps,  height=5.6cm} %6.3cm
\epsfig{file=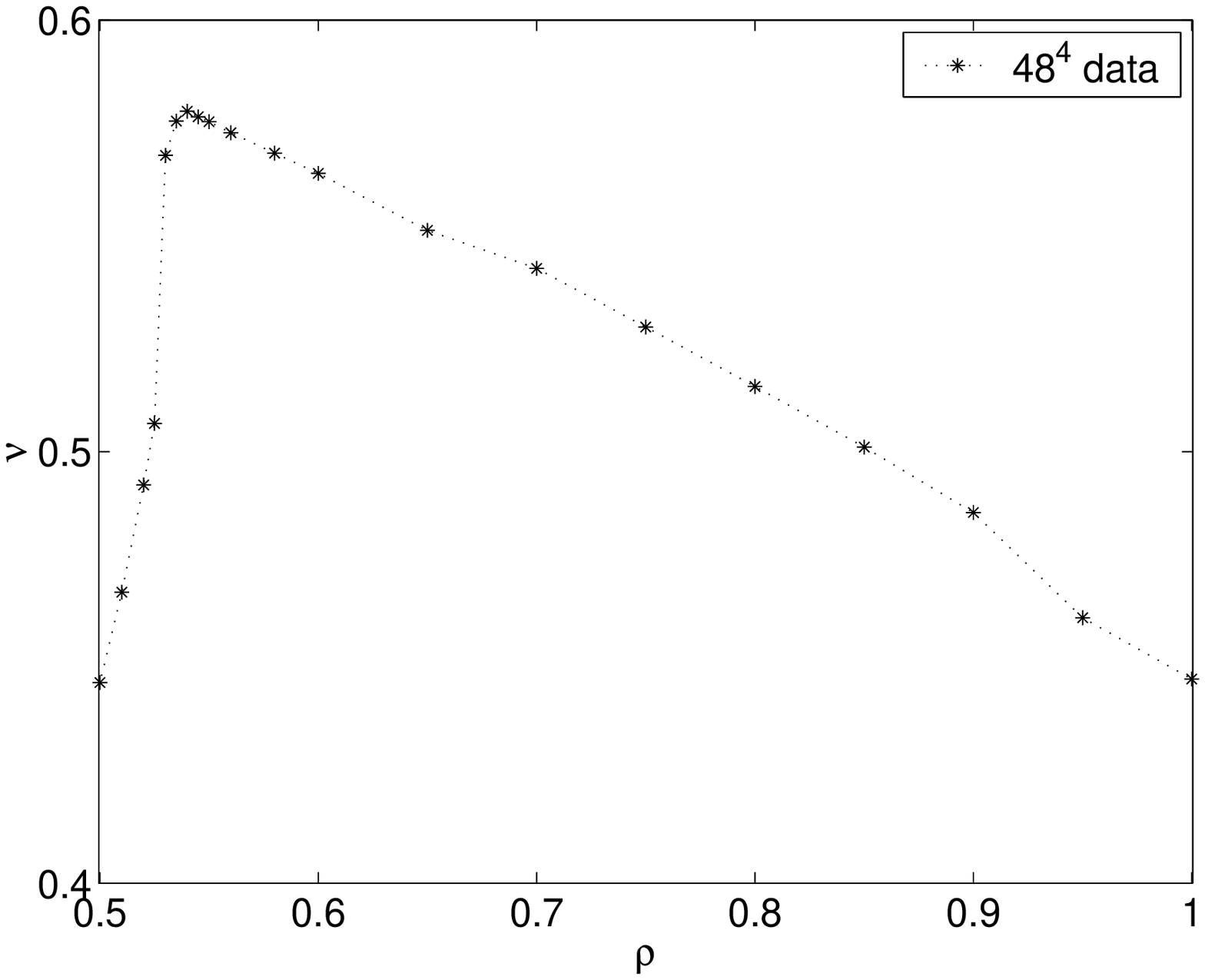,height=5.7cm} %6.4cm}
\vspace{-10mm}
\caption{Localization $\nu$ versus $\rh$ at $\be\!=\!6.0$ without filtering
and after 1 step of APE/HYP smoothing (left) and the same relation in the free
theory (right); here the optimum $\rh$ is significantly smaller than 1.}
\vspace{2mm}
\hspace{8mm}
\epsfig{file=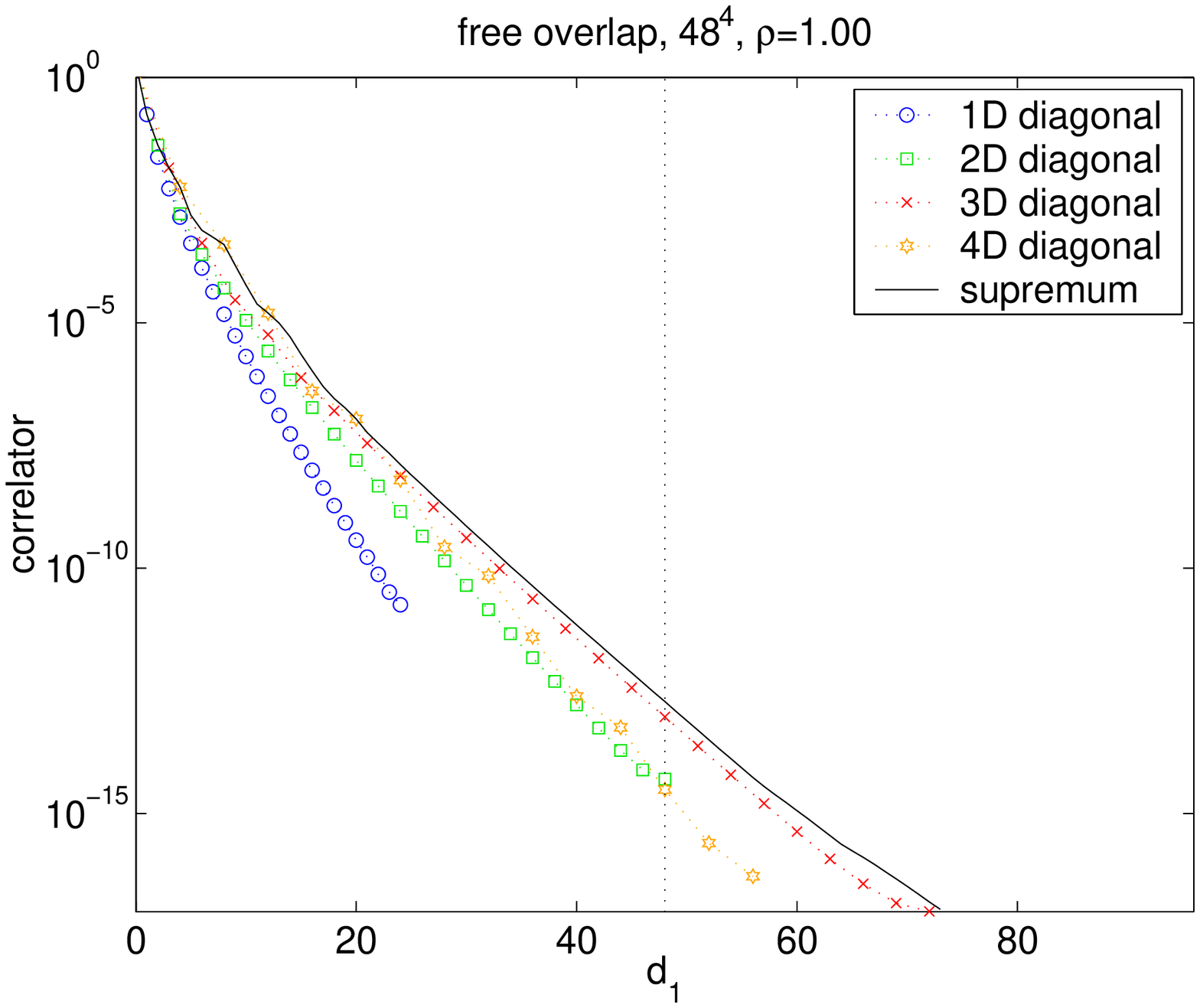,height=5.9cm} %6.6cm}
\epsfig{file=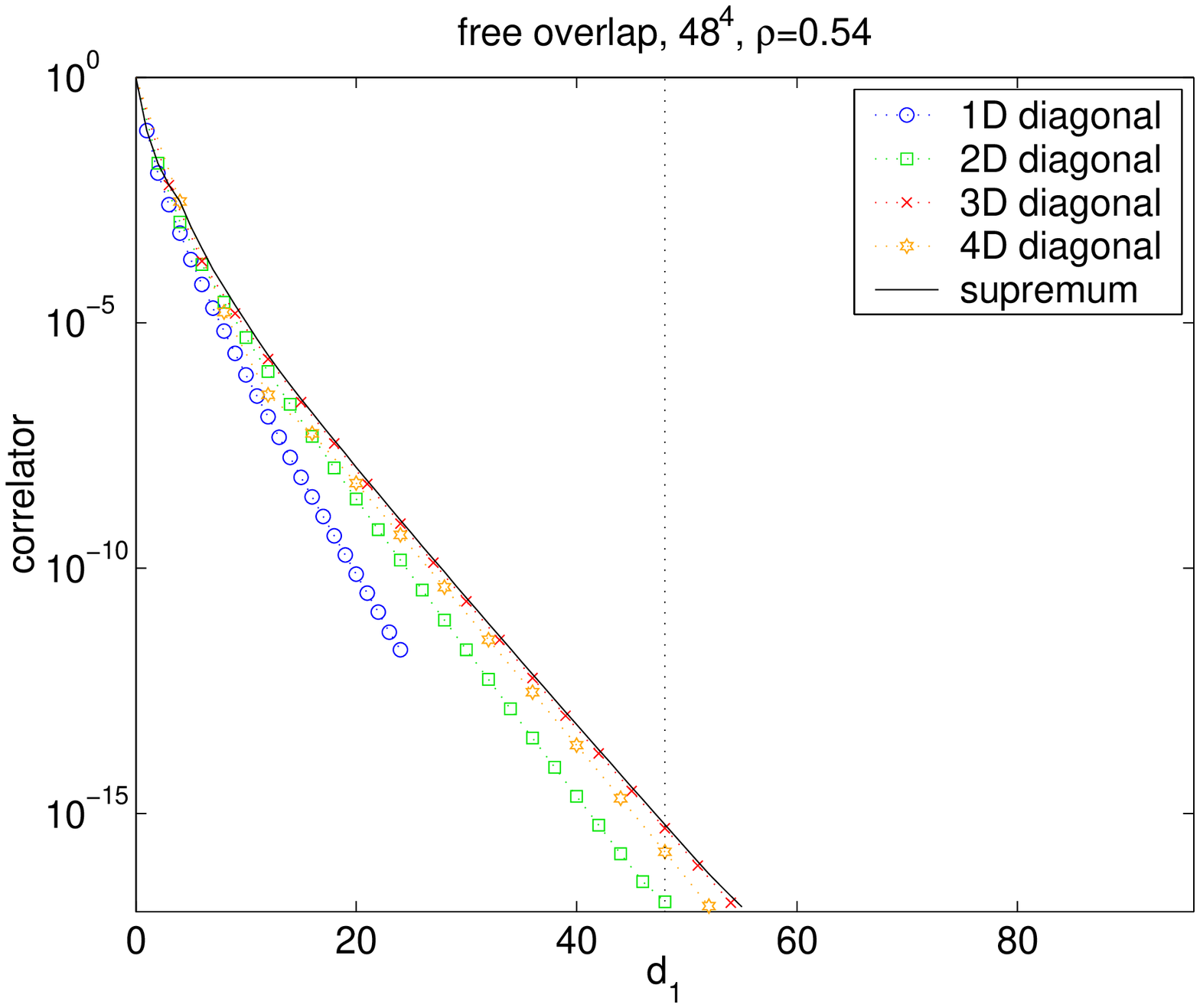,height=5.9cm} %6.6cm}
\vspace{-10mm}
\caption{Fall-off pattern (vs.\ ``taxi driver'' distance $d_1$) of the free
overlap operator on-axis (open circles) and for several directions (other
symbols) with supremum (full line) at $\rh=1$ (left) and $\rh=0.54$ (right).}
\vspace{4mm}
\hspace{8mm}
\epsfig{file=0506027.figs/za16_large.eps,height=5.0cm} %5.6cm}
\epsfig{file=zap.eps,                    height=5.0cm} %5.6cm}
\vspace{-10mm}
\caption{AWI mass versus bare mass %(both in lattice units)
at $\be\!=\!6.0$ and $\rh\!=\!1$ for several filtering recipes (left). The
quadratic fits are unconstrained and still go through zero, %(within errors),
indicating good chiral symmetry. The inverse slope $Z_A$ versus $\be$ (right)
shows a much milder dependence with filtering than without, even with
$\rh\!=\!1$ kept fixed.}
\end{figure*}

\subsection{Spectral properties of $|\HWr|$}\vspace{2mm}

We now focus on how the speedup of the overlap construction with UV-filtering
can be understood in terms of the spectrum of the underlying hermitean Wilson
operator $\HWr\!=\!\gaf(\DW\!-\!\rh)$.
We use the Wilson gauge action, and for technical details we refer to the
original publication \cite{Durr:2005an}.

The numerical effort to construct the massless operator (\ref{def_overzero}) is
in good approximation proportional to the condition number $1/\ep$ of $|\HWr|$.
Fig.\,1 shows $1/\ep$ on 25 quenched configurations at $\be\!=\!6.0$.
Evidently, the condition numbers get dramatically reduced through any type of
filtering.
In practice, overlap calculations use the projection trick, i.e.\ the subspace
pertinent to the lowest few eigenvalues is treated exactly
\cite{Hernandez:2000sb}.
The interesting news is that even after this trick has been applied, the
condition number in the orthogonal complement is improved through filtering
(right panel).
The data are for $\rh\!=\!1$, with $\rh\!>\!1$ the saving is slightly smaller.

It has been argued that the spectral density of $|\HWr|$ getting too
large will eventually make the overlap construction break down on coarse
lattices \cite{mobilityedge}.
Fig.\,2 shows the cumulative eigenvalue distribution (CED) of $|\HWr|$ with
$\rh\!=\!1$ on an extremely coarse ($\be=5.66$) and a fairly smooth
($\be\!=\!6.0$) lattice.
The slope at the origin is the quantity of interest, $\rh(0)$.
The filtering clearly reduces $\rh(0)$, but by no means does it reduce it to
zero.
A point worth noticing is that the unfiltered CED on the coarse lattice has
a rather different shape.
At strong coupling the radii of the Wilson spectrum are considerably smaller
than 1, and our choice to keep the shift parameter $\rh\!=\!1$ fixed lets
us ``loose'' the fermion.
In summary, the overlap construction breaks down on too coarse lattices, but
the good news is that $(i)$ one would notice from the CED and ($ii$) the
breakdown gets delayed through filtering ($\be\!=\!5.66$ seems fine if at
least 1 filtering step is applied).

Another question is what happens to the spectral density at weak coupling.
Fig.\,3 shows the CED of $|\HWr|$ in 2D in a log-log representation.
With filtering a much higher fraction of the eigenvalues is near $\la\!=\!1$,
and in this sense the picture is similar to what one would get if one were to
use an approximately chiral kernel -- in that case the ``mobility edge'' would
be at $\la\!=\!1$ and $\rh(0)\!=\!0$.
The high statistics study in 2D presented on the right suggests that $\rh(0)$
decays exponentially in $\be$ and that filtering continues to reduce $\rh(0)$
in the weak coupling regime.

\subsection{Localization properties of $\Dov$}\vspace{2mm}

It is evident from (\ref{def_overzero}) that $\Dov$ cannot be ultralocal, but
the question is whether it is local, i.e.\ whether the couplings in
$\Dov(x,y)$ would decay exponentially in $|x\!-\!y|$ (in some norm).
This is important to guarantee that LQCD with $\Dov$ is in the right
universality class.

We measure the localization function \cite{Hernandez:1998et}
\beq
f(d_1)=\mr{sup}\Big\{||(D_\mr{ov}\et)(x)||_2\;\Big|\;||x\!-\!0||_1\!=\!d_1\Big\}
\label{supremum}
\eeq
with $\et$ a normalized source vector at the origin.
Here, $d_1$ is the distance in the ``taxi driver'' metric.
The localization $\nu$ is the ``effective mass'' of $f(d_1)$ for a $d_1$ which
is sufficiently far from the maximal one ($2L$ in a $L^4$ box) to avoid finite
volume effects.
Fig.\,4 shows a scan of $\nu$ as a function of the shift parameter $\rh$ at
$\be\!=\!6.0$.
Without filtering the optimized $\rh$ is near $1.4$ \cite{Hernandez:1998et}.
After just 1 APE step the extremum shifts towards $\rh\!\simeq\!1$ and with
1 HYP step it is near $\rh\!\simeq\!0.8$.
This finding should not come as a surprise, since in the free theory a value
$\nu_\mr{max}\!\simeq\!0.58$ is realized for $\rh_\mr{opt}\!\simeq\!0.54$.
We consider this a strong rationale for \emph{not} tuning $\rh$ to an
``optimum'' value at some standard coupling but for staying with the
canonical choice $\rh\!=\!1$.
In passing we note that $\rh\!=\!1$ would also minimize the condition number
of $|\HWr|$ in the free field limit.

Our choice to use the ``taxi driver'' distance was motivated by the standards
in the literature \cite{Hernandez:1998et}.
Fig.\,5 shows $f(d_1)$ in the free theory for several directions (on-axis and
space-diagonals) together with the supremum.
One notices that the slope in the 2D-diagonal direction is roughly a factor
$\sqrt{2}$ smaller than on-axis, in the 3D-diagonal direction (the one which
dominates the supremum) it is about a factor $\sqrt{3}$.
In other words, if we had chosen the ``beeline'' metric $||.||_2$, then the
maximum localization in the free-field case would have turned out to be roughly
$0.58\sqrt{3}\!\simeq\!1$.

\subsection{Axialvector renormalization for $\Dov$}\vspace{2mm}

In phenomenological studies theoretical uncertainties of lattice-to-continuum
renormalization factors often limit the precision of the final answer.
In this respect it seems important that such factors would be close to 1
or (at least) show a mild dependence on the gauge coupling.

We determine $Z_A$ via the axial Ward identity (AWI).
Specifically, we use the ``chirally rotated'' quark fields
\cite{Capitani:1999uz} to construct the pseudoscalar density and the
(naive) axialvector current
\bea
P(x)\!&\!=\!&\!\psb_1(x)\gaf
\big[\big(1-{a\ovr2\rh}\Dov\big)\ps_2\big](x)
\label{def_p_rot}
\\
A_\mu(x)\!&\!=\!&\!\psb_1(x)\ga_\mu\gaf
\big[\big(1-{a\ovr2\rh}\Dov\big)\ps_2\big](x)
\label{def_a_rot}
\eea
where below the fields (flavors) $\ps_{1,2}$ will be taken as solutions to the
massive overlap operator
\beq
D_{\mr{ov},m}=\Big(1-{am\ovr2\rh}\Big)D_\mr{ov}+m
\label{def_overmass}
\eeq
with mass $m_{1,2}$, respectively.
The ``rotating'' operator \cite{Capitani:1999uz} in
(\ref{def_p_rot},\ref{def_a_rot}) is still the massless one.
With these at hand one forms the ratio
\beq
\rh(t,m_1,m_2)=
{\sum_\mb{x} \<P(\mb{x},t) P^c(\mb{0},0)\> \ovr
\sum_\mb{x} \<\bar\nab_{\!4}\!A_4(\mb{x},t) P^c(\mb{0},0)\>}
\label{def_awi}
\eeq
which yields the sum of the AWI quark masses,
\beq
\rh(t,m_1,m_2)\stackrel{t\to\infty}{\longrightarrow}
m_1^\mr{AWI}\!+\!m_2^\mr{AWI}+O(a^2)
\;.
\eeq

Fig.\,6 shows the plateau value $\rh(m_1\!+\!m_2)$ for a variety of $(m_1,m_2)$
combinations versus the sum $m_1\!+\!m_2$.
Since each $m_1\!+\!m_2$ is realized in various combinations and we see no
spread, we conclude that isospin is a perfect symmetry (in this ratio and
within our statistical precision).
Using
\bdm
\rh=
\mr{const}+{1\ovr Z_A}(m_1\!+\!m_2)+\mr{const}\,(m_1\!+\!m_2)^2
\edm
as our fit ansatz, we determine the offset at the origin and $Z_A$.
The former is consistent with zero, and this means that we have indeed good
chiral symmetry.
How the $Z_A$ factors would depend on the gauge coupling is shown on the
right.
Our data are for $\rh\!=\!1$ and stem from \cite{Durr:2005an,Durr:2005zk}.
The data for $\rh\!=\!1.4$ come from \cite{Wennekers:2005wa}.
This compilation shows that already a single HYP step at fixed $\rh\!=\!1$
manages to tame the $\be$-dependence of $Z_A$ in a more efficient way than
carefully tuning $\rh$ could possibly do.
In fact, the $Z_A$ values of the filtered overlap are so close to 1 that even
a perturbative evaluation at the 1-loop level might work fine (as was done,
for a different operator, in \cite{DeGrand:2002va}).

\subsection{Technical summary}\vspace{2mm}

Given the discussion in the two previous subsections it is clear that the
UV-filtering does not only serve the purpose of speeding up the overlap
construction, it really improves the physics properties of the resulting
$\Dov$.
Let us take the opportunity to highlight some technical points:
\begin{itemize}
\itemsep-2pt
\item[(a)] Filtering yields an $O(a^2)$-redefinition of the overlap
(at fixed $\rh,\al_\mr{smear},N_\mr{iter}$).
\item[(b)] The implementation effort is minimal; just evaluate $\Dov$ on
a smeared copy of the original gauge configuration.
\item[(c)] The idea is well-suited for dynamical simulations with HMC
\cite{dynover}, provided the stout-link recipe \cite{Morningstar:2003gk} is
used.
\item[(d)] The advantages are independent of how the $(.)^{-1/2}$ prescription
is implemented and naturally carry over to the domain-wall \cite{domainwall} or
Moebius \cite{Brower:2005qw} varieties.
\end{itemize}
Regarding point (a) it should be clear that there is no conflict with
the statement in \cite{indextheorem,Luscher:1998pq} that ``the'' overlap
yields a sound definition of ``the'' topological charge of a configuration $U$.
It is true that for some configurations in a given ensemble the filtered
overlap produces a mode excess number
\beq
q^\mr{HYP}={1\ovr2\rh}\trc(\gaf\,a\Dov^\mr{HYP}[U])
\eeq
which differs from the unfiltered version
\beq
q^\mr{none}={1\ovr2\rh}\trc(\gaf\,a\Dov^\mr{none}[U])
\;.
\eeq
It is important to notice that the same fact holds w.r.t.\ two unfiltered
overlap varieties with unequal $\rh$.
The statement in \cite{indextheorem,Luscher:1998pq} is that the number of
such ``ambiguous'' configurations dies out quickly enough with
$\be\!\to\!\infty$ to make the impact on any physical observable
(e.g.\ $\ch_\mr{top}$) an $O(a^2)$ effect.
These issues are discussed in \cite{oldfilt_over,DuHo_schwinger,Durr:2005an}.

%%%%%%%%%%%%%%%%%%%%%%%%%%%%%%%%%%%%%%%%%%%%%%%%%%%%%%%%%%%%%%%%%%%%%%%%%%%%%%%

\section{EXPLORATORY SCALING TEST}

\begin{figure}[!t]
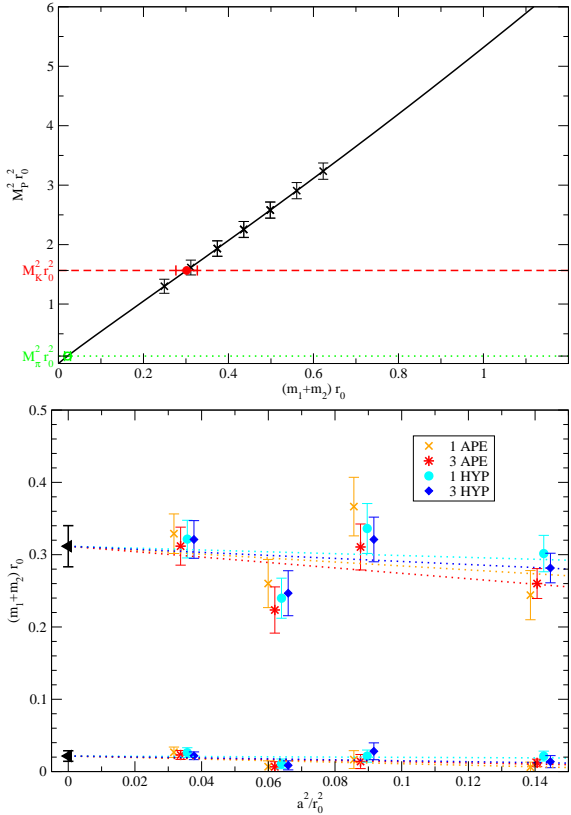

\epsfig{file=0508085.figs/mpi_hyp1_08.eps,width=7.5cm}
\epsfig{file=0508085.figs/scal_m.eps,     width=7.5cm}
\vspace{-14mm}
\caption{$M_P^2$ versus $m_1\!+\!m_2$ on our coarsest lattice (1\,HYP operator)
and the pertinent scaling plot for $2m_{ud}r_0$ and $(m_{ud}\!+\!m_s)r_0$ (all
filterings).}
\end{figure}

\begin{figure}[!t]
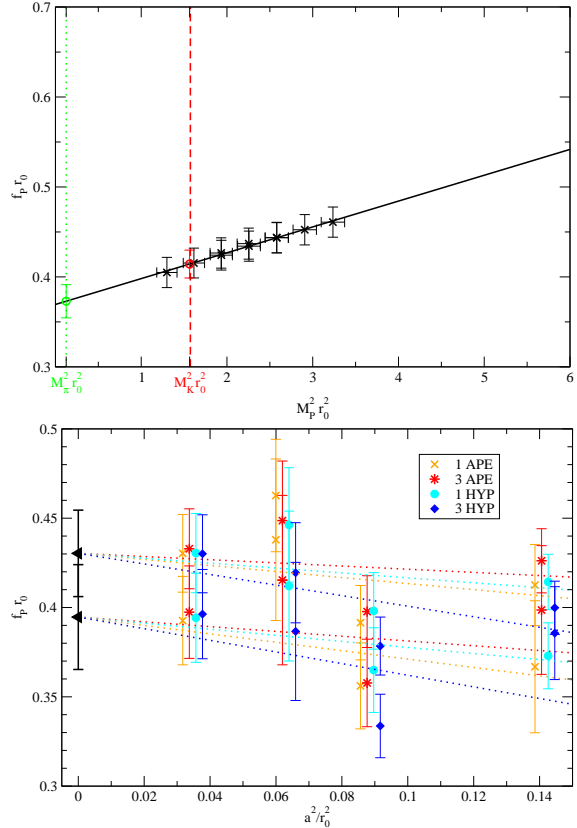

\epsfig{file=0508085.figs/fpi_hyp1_08.eps,width=7.5cm}
\epsfig{file=0508085.figs/scal_f.eps,     width=7.5cm}
\vspace{-14mm}
\caption{$f_P$ versus $M_P^2$ on our coarsest lattice (1\,HYP) and the
scaling plot for $f_\pi r_0$ and $f_K r_0$.}
\end{figure}

Given the promising localization properties of the filtered overlap, it is
natural to ask whether this will lead to a better scaling behavior.
There have been other scaling studies with the overlap action
\cite{Wennekers:2005wa,DuHo_schwinger,scaling_over}, but here we shall focus on
the first continuum result for $m_{ud},m_s$ and $f_\pi,f_K$ from quenched
overlap data \cite{Durr:2005ik}.

On the technical level, one more ingredient is needed, $Z_S\!=\!Z_m^{-1}$.
We compute it with the RI-MOM method \cite{Martinelli:1994ty}.
It turns out that the filtering leads to a rather mild $\be$-dependence of the
scalar renormalization factor, too \cite{Durr:2005ik}.

For the scaling study we use 4 couplings, $\be\!=\!5.66,5.76,5.84,6.0$,
with lattice dimensions such as to have one matched spatial box size
$L\!=\!1.5\fm$.
We choose 4 quark masses, ranging from about ${1\ovr3}m_s^\mr{phys}$ to
$m_s^\mr{phys}$, and 4 filterings (1\,APE, 3\,APE, 1\,HYP, 3\,HYP).
We do, however, not consider the unfiltered operator, simply because it is
too demanding for our computational resources.

Fig.\,7 shows the pseudoscalar mass squared versus the sum of the valence quark
masses (the Sommer radius $r_0$ makes them dimensionless) on our coarsest
lattice with the 1\,HYP version of $\Dov$.
The analogous plots at the other couplings look rather similar (not shown).
Imposing the physical value $M_Kr_0\!=\!1.251$ yields a definition of
$(m_s\!+\!m_{ud})r_0$ at this particular lattice spacing.
Likewise, imposing $M_\pi r_0\!=\!0.3537$ yields $2m_{ud}r_0$ at this coupling.
Repeating this procedure for the other operators and on the finer lattices,
we are finally in a position to extrapolate to the continuum.
We obtain the quenched continuum values
\bea
m_s^\mr{\overline{MS}}(2\GeV)\!&\!=\!&\!119(10)(7)\MeV
\\
m_s/m_{ud}\!&\!=\!&\!23.7(7.1)(4.5)
\eea
with the first error being statistical, the second systematic (up to
quenching).

Fig.\,8 shows the pseudoscalar decay constant versus the pseudoscalar mass
squared, again on our coarsest lattice with the 1\,HYP operator.
Superimposing the data obtained with other filterings or at weaker coupling
would reveal a rather good agreement.
Using the same values for $M_Kr_0$ and $M_\pi r_0$ as before, we get $f_Kr_0$
and $f_\pi r_0$ at this coupling.
Repeating this procedure on the finer lattices, we can extrapolate the
pseudoscalar decay constants to the continuum.
We obtain
\bea
f_K\!&\!=\!&\!170(10)(2)\MeV
\\
f_K/f_\pi\!&\!=\!&\!1.17(4)(2)
\eea
in the continuum with a systematic uncertainty that does not include quenching
effects.

Admittedly, our continuum extrapolations are somewhat courageous -- we start
from a lattice as coarse as $a^{-1}\!=\!1\GeV$ and have $a^{-1}\!=\!2\GeV$ on
the finest one.
Given this range and our statistical precision, we can only conjecture that we
are in the Symanzik scaling regime.
%Still, this statement remains true with any statistical precision and any
%range of couplings.
But in a strict sense this statement remains true with any statistical
precision and any range of couplings.

It is interesting to compare our continuum results to those of
\cite{Aoki:2002fd}.
The central values are consistent, but the relevant aspect is the error.
The CP-PACS result was thought to be the ``final'' quenched value and consumed
about half a year of runtime on their then new machine.
Our calculation used an equivalent of less than 1 year on a stand-alone PC.
And yet the net combined error (statistical and systematics, without
quenching) is almost the same.
This illustra\-tes that the absence of additive mass renormalization, due to
exact GW symmetry, has important practical consequences.

%%%%%%%%%%%%%%%%%%%%%%%%%%%%%%%%%%%%%%%%%%%%%%%%%%%%%%%%%%%%%%%%%%%%%%%%%%%%%%%

\section{SUMMARY}

The main features of overlap fermions with UV-filtering may be summarized as
follows.
\begin{enumerate}
\itemsep-2pt
\item
The overlap construction per se (with any undoubled kernel) brings a
variety of invaluable conceptual features like on-shell chiral symmetry and a
sound definition of the topological charge via the index theorem.
\item
Using an UV-filtered Wilson kernel several technical advantages can be
achieved: there is no need to tune the shift parameter $\rh$, the locality
of the resulting $\Dov$ and the normality of the kernel $\DWr$ are improved,
furthermore $\HWr$ has a better condition number, and renormalization
factors like $Z_A$ depend only weakly on the coupling.
\item
The details of the filtering procedure seem rather irrelevant; one may use
APE, HYP or stout-smearing with any set of parameters.
The crucial issue from a conceptual viewpoint is that these parameters
($\al_\mr{smear}$ and $N_\mr{iter}$) are kept fixed when $\be$ changes,
just like the other parameter $\rh$ must stay fixed.
In practice it seems advisable to stay with a moderate amount of
filtering, e.g.\ 1 to 3 iterations with standard parameters.
\item
Our exploratory scaling study suggests that such filtered overlap quarks
enjoy good scaling properties already on rather coarse ($a^{-1}\simeq1-1.5\GeV$)
lattices.
Under the proviso that more detailed studies with higher statistics confirm
this finding, it seems conceivable that overlap quarks become the method of
choice whenever high-precision results in the continuum are sought-for.
\end{enumerate}

{\bf Acknowledgments}: One of us (S.D.) likes to acknowledge useful discussions
with Gian Carlo Rossi and Tony Kennedy during the workshop.

%%%%%%%%%%%%%%%%%%%%%%%%%%%%%%%%%%%%%%%%%%%%%%%%%%%%%%%%%%%%%%%%%%%%%%%%%%%%%%%

%%%%%%%%%%%%%%%%%%%%%%%%%%%%%%%%%%%%%%%%%%%%%%%%%%%%%%%%%%%%%%%%%%%%%%%%%%%%%%%

\end{document}